\documentstyle[12pt]{article}
\begin{document}
\begin{titlepage}
\scrollmode

\begin{center}
{\Large\bf On Generalized Super-Coherent States}
\end{center}

\vspace{0.5cm}
\begin{center}
{\bf    M.~DAOUD,$^1$
     Y.~HASSOUNI,$^1$
and     M.~KIBLER$^2$}
\end{center}

\begin{center}
{$^1$Laboratoire de Physique Th\'eorique, Facult\'e des Sciences de Rabat}\\
{Universit\'e Mohammed V, avenue Ibn Batouta, B.P. 1014, Rabat, Morocco}
\end{center}

\begin{center}
{$^2$Institut de Physique Nucl\'eaire de Lyon, IN2P3-CNRS et UCBL}\\
{43 boulevard du 11 novembre 1918, F-69622 Villeurbanne Cedex, France}
\end{center}

\vspace{2cm}
\begin{abstract}

A set of operators, the so-called $k$-fermion operators, that interpolate
between boson and  fermion  operators are introduced through the 
consideration of an algebra arising from two non-commuting quon algebras. 
The deformation parameters 
$q$ and  $1/q$  for these quon algebras are roots of unity with 
$q = {\rm exp}( { 2 \pi {\rm i} / k } )$ and 
$k \in {\bf N} \setminus \{ 0,1 \}$. 
The case $k=2$ corresponds to fermions and
the limiting case $k \to \infty$ to bosons. 
Generalized coherent states
(connected to $k$-fermionic states) and super-coherent states 
(involving a  $k$-fermionic sector and a purely bosonic sector) are 
investigated. 

\end{abstract}

\vskip 6.5 true cm
\noindent 
Paper to appear in the Proceedings of the {\bf VIII International Conference on
Symmetry Methods in Physics} (JINR, Dubna, Russia, 28 July -- 2 August 1997). 
The Proceedings of the Conference will be published in the Russian Journal of
Nuclear Physics (Yadernaya Fizika). 

\vfill
\thispagestyle{empty}
\end{titlepage}

\newpage

%\baselineskip 0.8 true cm

\section{Introduction}

The interest of $q$-deformations for statistical physics is still 
very high in the community of physicists and mathematicians. In recent years, 
many works have been devoted to statistics of $q$-bosons, $q$-fermions 
and quons (see, for instance, Ref.~[1] and references therein). 
This paper is devoted to $k$-fermions 
which are objects interpolating between fermions (corresponding to $k=2$) and 
bosons   (corresponding to $k \to \infty$).

The material in the present paper is organized as follows. 
 We first discuss (in Section 2) the $k$-fermionic 
algebra $\Sigma_q$, 
where $q := {\rm exp} (2 \pi {\rm i} / k)$ 
with $k \in {\bf N} \setminus \{ 0,1 \}$, in terms of 
generalized Grassmann variables. Then, we introduce (in Section 3) 
generalized coherent 
states. Finally, the notion of fractional super-coherent states is introduced 
(in Section 4) from a certain limit
of the well-known deformed coherent states. 

\section{The $k$-fermions} 

\subsection{The $k$-fermionic algebra $\Sigma_q$}

We first introduce the $k$-fermionic algebra $\Sigma_q$. The algebra $\Sigma_q$
is generated by five operators $a_+$, 
                               $a_-$, $a_+^+$, 
                                      $a_-^+$ and $N$. We assume that $N$ is an
Hermitean operator, that $a_+^+$ (respectively, $a_-^+$) is the adjoint of 
$a_+$ (respectively, $a_-$) and that these operators satisfy
$$
a_- a_+ - q a_+ a_- = 1 
\iff 
a_+^+ a_-^+ - {\bar q} a_-^+ a_+^+ = 1
\eqno (1{\rm a}) 
$$
$$
N a_+ - a_+ N = + a_+
\iff 
N a_+^+ - a_+^+ N = - a_+^+
\eqno (1{\rm b}) 
$$
$$
N a_- - a_- N = - a_-
\iff 
N a_-^+ - a_-^+ N = + a_-^+
\eqno (1{\rm c})
$$
$$
(a_+)^k = (a_-)^k = 0
\iff
(a_+^+)^k = (a_-^+)^k = 0
\eqno (1{\rm d})
$$
$$
a_- a_+^+ = {\bar q}^{{1 \over 2}} a_+^+ a_- 
\iff 
a_+ a_-^+ = {     q}^{{1 \over 2}} a_-^+ a_+
\eqno (1{\rm e})
$$
where the complex number  
$$
q := {\rm exp} \left( {2 \pi {\rm i} \over k} \right) 
\quad {\rm with} \quad k \in {\bf N} \setminus \{ 0,1 \}
$$
is a root of unity. (In Eq.~(1), ${\bar q}$ stands for 
the complex conjugate of $q$.) The algebra $\Sigma_q$ clearly involves 
two non-commuting quon algebras 
$A_  q $     (spanned by $a_+$,   $a_-$   and $N$) and 
$A_{\bar q}$ (spanned by $a_-^+$, $a_+^+$ and $N$). 

In view of the defining relations (1), the operators  $a_+$, 
                               $a_-$, $a_+^+$, 
                                      $a_-^+$ and $N$ act 
on a Fock space 
$ {\cal F} := \{ | n \rangle : n = 0, 1, \cdots, k-1 \} $ 
with card~${\cal F}=k$. 
Furthermore, we chose a representation of $\Sigma_q$ in the
following way. The action of $N$ is standard in the sense that  
$$
N | n \rangle = n | n \rangle 
$$
while the action of the remaining operators is given by 
$$
a_- | n   \rangle = 
\left( \left[ n     \right]_q \right)^{{1 \over2}} 
    | n-1 \rangle \quad {\hbox{with}} \quad a_- |     0 \rangle = 0
$$
$$
a_+^+ | n \rangle = 
\left( \left[ n \right]_{\bar q} \right)^{1 \over 2} | n-1 \rangle
\quad {\hbox{with}} \quad a_+^+ |   0 \rangle = 0
$$
and 
$$
a_+ | n   \rangle = 
\left( \left[ n + 1 \right]_q \right)^{{1 \over2}} 
    | n+1 \rangle \quad {\hbox{with}} \quad a_+ | k - 1 \rangle = 0
$$
$$
a_-^+ | n \rangle = 
\left( \left[ n + 1 \right]_{\bar q} \right)^{1 \over 2} | n+1 \rangle
\quad {\hbox{with}} \quad a_-^+ | k-1 \rangle = 0
$$
where  the symbol $[~~]_q$ is defined by 
$$
[X]_q := {1 - q^X \over 1 - q}
$$
for any  operator or  number $X$. 
 Thus, the operators
 $a_-$ and $a_+^+$ behave like annihilation operators, the operators
 $a_+$ and $a_-^+$        like creation     operators and 
the operator $N$ like a number operator. 

 The state vector $ | n \rangle $ can be written as
$$
| n \rangle = { (a_+)^n \over ([n]_q          !)^{1 \over 2} } \> |0 \rangle 
\quad
{\hbox{or}} 
\quad 
| n \rangle = { (a_-^+)^n \over ([n]_{\bar q} !)^{1 \over 2} } \> |0 \rangle 
\quad
{\hbox{for}} 
\quad 
n = 0, 1, \cdots, k-1
$$
where, as usual, the $p$-deformed factorial $[n]_p$
is defined by (with $p=q$ and ${\bar q}$)
$$
\lbrack n \rbrack_p ! := 
\lbrack 1 \rbrack_p 
\lbrack 2 \rbrack_p \cdots 
\lbrack n \rbrack_p \quad {\hbox{for}} \quad 
n \in {\bf N} \setminus \lbrace 0 \rbrace \quad {\hbox{and}} \quad
\lbrack 0 \rbrack_p ! := 1
$$

In the specific case $k=2$, the algebra $\Sigma_{-1}$ corresponds to ordinary 
fermion operators with $a_+^+ = a_-$ and $a_-^+ = a_+$ for which we have 
$(a_-)^2 = (a_+)^2 = 0$, 
a relation that reflects the Pauli exclusion principle. 
In the limiting case 
$k \to \infty$, the algebra $\Sigma_{+1}$ corresponds to ordinary  
boson   operators  with $a_+^+ = a_-$ and $a_-^+ = a_+$. 
For $k$ arbitrary, the 
algebra $\Sigma_q$ corresponds to $k$-fermion operators 
$a_-$  and  $a_+$ 
(with their adjoint $a_-^+$ and $a_+^+$, respectively) 
that interpolate between fermion and boson operators~; 
the space ${\cal F}$ is of dimension $k$ for the $k$-fermionic algebra 
$\Sigma_{q}$ (i.e., two-dimensional for the fermionic algebra $\Sigma_{-1}$
and infinite-dimensional            for the bosonic   algebra $\Sigma_{+1}$). 

\subsection{Grassmannian realization of $\Sigma_q$} 

We give here some preliminaries useful for obtaining a Grassmannian realization
of the algebra $\Sigma_q$. Equation 
(1d) suggests that we use generalized Grassmann variables 
(see Refs.~[2-5]) $z$ and ${\bar z}$ 
such that  
$$
z^k = {\bar z}^k = 0 
\eqno (2)
$$
(The particular case $k = 2$ corresponds to ordinary Grassmann variables.) We 
then introduce the $\partial_{z}$- and $\partial_{\bar z}$-derivatives via 
$$
\partial_z f(z) := {f(qz) - f(z) \over (q - 1) z}, 
\qquad
\partial_{\bar z} g({\bar z}) := 
{g({\bar q} {\bar z}) - g({\bar z}) \over ({\bar q} - 1) {\bar z}}
\eqno (3)
$$
where $f :       z  \mapsto f(      z )$ 
  and $g : {\bar z} \mapsto g({\bar z})$ 
are arbitrary functions. 
The linear operators $\partial_{z}$ and $\partial_{\bar z}$ satisfy 
$$
\partial_z z^n = [n]_q \> z^{n-1}, 
\qquad 
\partial_{\bar z} {\bar z}^n = [n]_{\bar q} \> {\bar z}^{n-1}
$$
for $n = 0, 1, \cdots, k-1$. Therefore, when 
$f(      z )$ and 
$g({\bar z})$ can be developed as
$$
f(z) = \sum_{n=0}^{k-1} a_n z^n,
\qquad
g({\bar z}) = \sum_{n=0}^{k-1} b_n {\bar z}^n
$$
where the coefficients $a_n$ and $b_n$ in the expansions are complex numbers,
we check that 
$$
(\partial_z)^k f(z) = 
(\partial_{\bar z})^k g({\bar z}) = 0
$$
Consequently, we shall assume that the conditions 
$$
(\partial_z)^k = (\partial_{\bar z})^k = 0
\eqno (4)
$$
hold in addition to Eq.~(2). 

From Eqs.~(2) and (4), the correspondences 
$$
a_-   \rightarrow \partial_z,        \qquad a_+   \rightarrow       z, 
                                     \qquad
a_+^+ \rightarrow \partial_{\bar z}, \qquad a_-^+ \rightarrow {\bar z}
\eqno (5) 
$$
clearly provide us with a realization of Eqs.~(1a) and (1d). 
Note that Eq.~(1e) leads to
$$
\partial_z \partial_{\bar z} = 
             {\bar q}^{{1 \over 2}} \> \partial_{\bar z}\partial_z, 
\qquad
z {\bar z} = {     q}^{{1 \over 2}} \> {\bar z} z
$$
in the realization based on Eq.~(5). 

\section{Generalized coherent states} 

There exists several methods for introducing coherent states. We can use the
action of a displacement operator on a reference state [6] 
or the construction of an eigenstate for an annihilation operator [7,8] 
or the minimisation of uncertainty relations [9]. In the case of the 
ordinary harmonic oscillator, the three methods lead to the same result (when
the reference state is the vacuum state). Here, the situation is a little bit
more intricate (as far as the equivalence of the three methods is concerned) 
and we chose to define the generalized 
                           coherent states or $k$-fermionic coherent states 
$|       z )$ and 
$| {\bar z})$ as follows
$$
|z)        := \sum_{n=0}^{k-1} {       z ^n \over ([n]_q       !)^{1 \over 2} } 
\> | n \rangle, 
\qquad
|{\bar z}) := \sum_{n=0}^{k-1} { {\bar z}^n \over ([n]_{\bar q}!)^{1 \over 2} }
\> | n \rangle
$$ 
where $z$ and $\bar z$ are generalized Grassmann variables that satisfy
Eq.~(2). It can be easily checked that 
the state vectors $| z )$ and $| {\bar z} )$ are eigenvectors of the operators
$a_-$ and $a_+^+$, respectively. More precisely, we have
$$
a_- |z) = z | z),
\qquad
a_+^+ | {\bar z} ) = {\bar z} | {\bar z} )
$$
The case $k=2$ corresponds to fermionic coherent states while the limiting case
$k \to \infty$ to bosonic coherent states.

We define 
$$
( z | := \sum_{n=0}^{k-1} \> \langle n | \>
{ {\bar z}^n \over ([n]_{\bar q}!)^{1 \over 2} },
\qquad
( {\bar z} | := \sum_{n=0}^{k-1} \> \langle n | \>
{ z^n \over ([n]_q!)^{1 \over 2} } 
$$
Then, the `scalar products' $( z' | z )$ and $( {\bar z}' | {\bar z} )$ 
follow  from  the ordinary scalar product 
$ \langle n' | n \rangle = \delta(n',n)$. For instance, we get 
$$
(z' | z) = \sum_{n=0}^{k-1} 
              { {\bar {z'}}^n z^n \over ([n]_{\bar q}! [n]_q!)^{1 \over 2} }
$$
In view of the relationship 
$$
[n]_{\bar q}!  = q^{- {1 \over 2} n(n-1)} \> [n]_q !
$$
and of the property 
$$
{\bar z}^n z^n = q^{- {1 \over 4} n(n-1)} \> ({\bar z} z)^n
$$
we obtain the following result 
$$
(z | z) = \sum_{n=0}^{k-1} { ({\bar z} z)^n \over [n]_q! }
\eqno (6)
$$
Similarly, we have 
$$
({\bar z} | {\bar z}) = \sum_{n=0}^{k-1} { (z {\bar z})^n \over [n]_{\bar q}! }
\eqno (7)
$$
By defining the $q$-deformed exponential ${\rm e}_q$ by 
$$
{\rm e}_q : x \mapsto {\rm e}_q (x) := \sum_{n=0}^{k-1} { x^n \over [n]_q! }
$$
we can rewrite Eqs.~(6) and (7) as 
$$
(       z |      z  ) = {\rm e}_      q  ({\bar z} z), \qquad 
( {\bar z}|{\bar z} ) = {\rm e}_{\bar q} (z {\bar z})
$$
(Observe that the summation 
in the exponential ${\rm e}_q$ is finite, for $k$ finite, rather than 
infinite as is usually the case in $q$-deformed exponentials.)

We guess that the $k$-fermionic coherent states $| z )$ and $| {\bar z} )$ 
form overcomplete sets with respect to some integration process 
accompanying the derivation process inherent to Eq.~(3). 
Following Majid and Rodr\'\i guez-Plaza [5], we consider the integration
process defined by
$$
\int dz \> z^p     = \int d{\bar z} \> {\bar z}^p := 0 \quad {\hbox{for}} 
\quad p = 0, 1, \cdots, k-2
\eqno (8{\rm a}) 
$$
and
$$
\int dz \> z^{k-1} = \int d{\bar z} \> {\bar z}^{k-1} := 1
\eqno (8{\rm b})
$$
Clearly, the integrals in (8) generalize the Berezin integrals corresponding
to $k = 2$. In the case where $k$ is arbitrary, we can derive the
overcompleteness property 
$$
\int d      z  \> |       z  ) \> 
\mu(z, {\bar z}) \> (       z  | \> d{\bar z} = 
\int d{\bar z} \> | {\bar z} ) \> 
\mu({\bar z}, z) \> ( {\bar z} | \> d      z  = 1
$$
where the function $\mu$ defined through
$$
\mu(z, {\bar z}) := \sum_{n=0}^{k-1} \> ( [n_q]! [n_{\bar q}]! )^{1 \over 2} \>
z^{k-1-n} \> {\bar z}^{k-1-n}
$$
may be regarded as a measure.

\section{Fractional super-coherent states} 

We now switch to $Q$-deformed coherent states of the type 
$$
|Z) := \sum_{n=0}^{\infty} { Z^n \over ([n]_Q !)^{1 \over 2} } \> | n \rangle
\eqno (9) 
$$
associated to a quon algebra $A_Q$ where $Q \in {\bf C} \setminus {S}^1$.
The latter states are simple deformations of the bosonic coherent states
(cf.~Ref.~[10]). The coherent state $| Z )$ may be considered 
to be an eigenstate, with the eigenvalue $Z \in {\bf C}$, of an annihilation
operator $b_-$ in a representation such that the operator $b_-$ and 
the associated creation operator $b_+$ satisfy
$$
b_- | n \rangle = \left( \left[ n \right]_Q \right)^{1 \over 2} 
| n - 1 \rangle \quad {\hbox{with}} \quad b_- | 0 \rangle = 0
$$
$$
b_+ | n \rangle = \left( \left[ n + 1 \right]_Q \right)^{1 \over 2} 
| n + 1 \rangle
$$
with $n \in {\bf N}$.  

For $Q \to q$, we have $[k]_Q! \to 0$. Therefore, the term 
$Z^k / ( [k]_Q! )^{1 \over 2}$ 
in Eq.~(9) makes sense for $Q \to q$ only if $Z \to z$,
where $z$ is a generalized 
Grassmann variable with $z^k = 0$. This type of reasoning has
been invoked for the first time in Ref.~[11]. (In [11], 
the authors show that there is an isomorphism between the braided line and the
one-dimensional super-space.)

It is the aim of this section to determine the limit 
$$
| \xi ) := \lim_{Q \to q} \lim_{Z \to z} \> |Z)
$$
when $Q$ goes to the root of unity $q = {\rm exp} (2 \pi {\rm i} / k)$ 
and $Z$ to a Grassmann
variable $z$. The starting point is to rewrite Eq.~(9) as 
$$
|Z) = \sum_{r=0}^{\infty} \sum_{s=0}^{k-1} 
{ Z^{rk + s} \over ( [rk + s]_Q! )^{1 \over 2} } \> | rk + s \rangle 
$$
Then, by making use of the formulas
$$
{ [k]_Q \over [r k]_Q } \to {1 \over r} \quad {\hbox{for}} \quad 
Q \to q \quad {\hbox{with}} \quad r \ne 0
$$
and
$$
{ [s]_Q \over [rk + s]_Q } \to 1 \quad {\hbox{for}} \quad 
Q \to q \quad {\hbox{with}} \quad s = 0, 1, \cdots, k-1
$$
we find that 
$$
\lim_{Q \to q} \lim_{Z \to z} 
{ Z^{rk + s} \over \left( [rk + s]_Q ! \right)^{1 \over 2} } 
= { z^s      \over ([s_q]!)^{1 \over 2} } \> 
  { \alpha^r \over (r    !)^{1 \over 2} }
\eqno (10)
$$
works for $s = 0, 1, \cdots, k-1$ and $r \in {\bf N}$. 
The complex variable $\alpha$ in Eq.~(10) is defined by 
$$
\alpha := 
\lim_{Q \to q} \lim_{Z \to z} { Z^k \over \left( [k]_Q! \right)^{1 \over 2} }
$$
Therefore, we obtain
$$
| \xi ) = \sum_{r=0}^{\infty} \sum_{s=0}^{k-1} 
{ z^s      \over \left( [s]_q! \right)^{1 \over 2} } 
{ \alpha^r \over (r!)^{1 \over 2} } \> | rk+s \rangle
$$
Finally, by employing the symbolic notation
$$
| rk + s \rangle \equiv | r \rangle \otimes | s \rangle
$$
we arrive at the formal expression 
$$
| \xi ) = 
\sum_{r=0}^{\infty} { \alpha^r \over (r!)^{1 \over 2} } \> | r \rangle 
\bigotimes 
\sum_{s=0}^{k-1}    { z^s \over \left( [s]_q! \right)^{1 \over 2} } \> 
                                                           | s \rangle
\eqno (11)
$$

We thus end up with the product of 
a       bosonic coherent state by 
a $k$-fermionic coherent state. This product shall be called a fractional 
super-coherent state. In the particular case $k=2$, it reduces to the product
of a    bosonic coherent  state by 
   a  fermionic coherent  state, i.e., to the 
super-coherent state associated to 
a super-oscillator [12]. In the
framework of field theory, Eq.~(11) means that in the limit $Q \to q$, every
field $\psi$ with values $\psi(Z)$ is transformed into a fractional super-field
$\Psi$ with value $\Psi(z, \alpha)$, $z$ being a generalized Grassmann variable
and $\alpha$ a bosonic variable. 

\section{Concluding remarks}

As a main result, the $k$-fermions introduced in the present 
paper can be ranged between fermions (for $k=2$) and bosons 
(for $k \to \infty$). This result is further emphasized by 
calculating the coherence factor $g^{(m)}$ for an assembly 
of $k$-fermions: We find that $g^{(m)} = 0$ for $m > k-1$ 
so that, in a many-particle scheme, a given state of
fractional spin $S= \frac{1}{k}$ cannot be occupied by
more than $k-1$ identical $k$-fermions. The $k$-fermions 
thus satisfy a generalized  Pauli  exclusion principle.

We close this paper by mentioning two open
questions. First, does the $W_{\infty}$  
algebra described by Fairlie, Fletcher and 
Zachos [13] plays an important role in the
symmetries inherent to $k$-fermions (see also
Ref.~[14])~? Second, what is the connection between $k$-fermions
and fractional super-symmetry for anyons [15,16], especially the anyons  
constructed from unitary representations of the group diffeomorphisms of the
plane [16]~? These matters should be the object of future works. 

\vskip 1 true cm 

\noindent
{\bf Acknowledgments}

\vskip 0.5 true cm 

One of the authors (M.K.) would like to thank 
W.S.~Chung, G.A.~Goldin and S.~Mashkevich 
for interesting comments on this work 
on the occasion of its presentation 
at the {\em VIII International Conference on
Symmetry Methods in Physics} in Dubna.


\begin{thebibliography}{99}
\itemsep=-3pt

\bibitem{1.} 
  M. Kibler and M. Daoud, pre-print physics/9712034. 

\bibitem{2.} 
  V.A. Rubakov and V.P. Spiridonov, 
  {\it Mod. Phys. Lett.}, Ser. A, 1988, vol. 3, p. 1337. 
 
\bibitem{3.}  
 A.T. Filippov, A.P. Isaev and A.B. Kurdikov, {\it Mod. Phys. Lett.}, Ser. A, 
 1992, vol. 7, p. 2129~; 
 {\it Int. J. Mod. Phys.}, Ser. A, 1993, vol. 8, p. 4973. 

\bibitem{4.}
  A. Le Clair and C. Vafa, {\it Nucl. Phys.}, Ser. B, 1993, vol. 401, p. 413. 

\bibitem{5.}  
 S. Majid and M.J. Rodr\'\i guez-Plaza, 
 {\it J. Math. Phys.}, 1994, vol. 35, p. 3753. 
    
\bibitem{6.}  
 A.M. Perelomov, {\it Generalized Coherent States and Their Applications} 
 (Springer, Berlin, 1986). 

\bibitem{7.}  
 R.J. Glauber, {\it Phys. Rev.}, 1963, vol. 131, p. 2766. 
 
\bibitem{8.}  
 E.C.G. Sudarshan, {\it Phys. Rev. Lett.}, 1963, vol. 10, p. 84. 
 
\bibitem{9.}  
 M.M. Nieto and L.M. Simmons, Jr., {\it Phys. Rev. Lett.}, 
 1978, vol. 41, p. 207. 
 
\bibitem{10.}  
 M. Arik and D.D. Coon, {\it J. Math. Phys.}, 1976, vol. 17, p. 524. 

\bibitem{11.}  
 R.S. Dunne, A.J. Macfarlane, J.A. de Azc\'arraga and J.C. P\'erez Bueno,
 {\it Phys. Lett.}, Ser. B, 1996, vol. 387, p. 294~; 
 {\it Int. J. Mod. Phys.}, Ser. A, 1997, vol. 12, p. 3275. 
 
\bibitem{12.}  
 Y. B\'erub\'e-Lauzi\`ere and V. Hussin, {\it J. Phys.}, Ser. A, 1993, vol. 26, 
 p. 6271. 

\bibitem{13} D.B. Fairlie, P. Fletcher and C.K. Zachos, {\it J. Math. Phys.},
1990, vol. 31, p. 1088. 

\bibitem{14} M. Daoud, Y. Hassouni and M. Kibler, 
in {\it Symmetries in Science X}, 
eds. B. Gruber and M. Ramek (Plenum Press, New York, 1998). 

\bibitem{15} J.M. Leinaas and J. Myrheim, {\it Nuovo Cimento}, Ser. B, 1977,
vol. 37, p. 1. 

\bibitem{16} G.A. Goldin, R. Menikoff and D.H. Sharp, {\it J. Math. Phys.},
1980, vol. 21, p. 650~; 1981, vol. 22, p. 1664~; G.A. Goldin and D.H. Sharp, 
{\it Phys. Rev. Lett.}, 1996, vol. 76, p. 1183.

\end{thebibliography}
\end{document}